\pdfoutput=1
\documentclass[12pt]{iopart}

\begin{document}

\title[]{Analytical Forms and Degeneracy of Quasinormal Modes for Kerr-Newman-de Sitter Black Holes}

\author{Zhong-Heng Li}

\address{Black Hole and Gravitational Wave Group,
 Zhejiang University of Technology,
 Hangzhou 310023, China}
\ead{zhli@zjut.edu.cn}
\vspace{10pt}
%\begin{indented}
%\item[]August 2017 (minor update March 2024)
%\end{indented}

\begin{abstract}
This study investigates the quasinormal modes of Kerr-Newman-de Sitter black holes for massless spin particles using the unified equation.
We derive analytical expressions for both the quasinormal mode frequencies and the radial wave functions. The frequencies are determined exclusively by the black hole parameters and the quantum numbers $n$ and $m$, while the radial wave functions also depend on the quantum number $k$, indicating a degeneracy in frequency. For identical quantum numbers, the frequency expression and the degree of degeneracy are the same for all massless spin particles, regardless of their specific properties. This implies that, through the observation of quasinormal modes, one can not only determine the black hole's parameters but also observe the phenomenon in which one type of particle reproduces the quasinormal mode of another. Our work thus provides a theoretical foundation for understanding this mimicking behavior as well.
\end{abstract}

%\maketitle

\section{Introduction}\label{sec:1}
Quasinormal modes are solutions to the wave equation subject to specific boundary conditions at the black hole horizon and far from the black hole (or cosmological horizon); they can also be defined as the poles of the black-hole Green's function for the Laplace transformed solutions [1]. The study of quasinormal modes and their frequencies for various black holes, spurred by Vishveshwara's initial discovery in 1970 [2], has since expanded greatly, as reflected in the many excellent reviews on the topic [1, 3-7].

The first detection of merging binary black holes by LIGO in 2015 [8]--including the ringdown phase--stimulated advanced research on quasinormal modes. This research pursued higher precision [9], investigated spectral stability [10], and explored deep connections with other physical fields [11-14], as these modes uniquely characterize the dominant component of the gravitational-wave ringdown of post-merger black holes and allow for precise tests of general relativity.

In this paper, we study the quasinormal modes of Kerr-Newman-de Sitter spacetime for massless spin particles. Our work differs from previous studies in that we derive analytical expressions for the quasinormal modes and discuss both the degeneracy of the quasinormal mode frequencies and the wave analogies for massless spin particles. Despite extensive study of quasinormal modes, analytic solutions remain exceedingly rare, as the governing equations are typically too complex to solve exactly. Even for the simplest Schwarzschild black hole, the Regge-Wheeler equation yields no analytic expression for its quasinormal mode spectrum. Since investigating degeneracy requires global information about the quasinormal mode solutions, there are almost no existing studies addressing the degeneracy of quasinormal mode frequencies. Furthermore, analogy studies of massless particles are of great importance, as they are closely related to fundamental questions such as whether gravitational waves can be simulated using electromagnetic waves in terrestrial laboratory settings.

The use of analogy is grounded in the physical principle that disparate systems can be governed by identical dynamical equations. This type of similarity analysis fosters the exchange and cross-fertilization of ideas across various scientific fields, with the concurrent development of gravitational and electromagnetic theory serving as a classic example. From the early ``gravitomagnetic'' explanation for Mercury's perihelion precession by Hozm$\ddot{u}$ller [15] and Tisserand [16, 17] to the recent visualization of binary black hole mergers by Boyeneni et al. [18]--who recast Einstein's equations into a form highly analogous to electrodynamics--analogy has served as a powerful tool for tackling complex problems in gravitational physics.

\section{Spacetimes and Wave equations}\label{sec:2}

The Teukolsky master equation [19, 20] has constituted a cornerstone for modeling gravitational waveforms from black hole mergers over recent decades. In light of the prospective high-precision observations by future space-based detectors (e.g., LISA 21,22], Taiji [23], and TianQin [24]), however, substantial consideration must now be given to including the effects of black hole charge and the cosmological constant. Consequently, the Kerr-Newman-de Sitter black hole has emerged as a particularly suitable and significant candidate. The form of the Kerr-Newman-de Sitter metric in Boyer-Lindquist-type coordinates is as follows [25, 26]:
\begin{eqnarray}\label{1}
ds^{2}=&\frac{\rho\bar{\rho}\Delta_{r}}{\Xi^{2}}(dt-a\sin^{2}\theta d\varphi)^{2} \nonumber\\
&-\frac{\rho\bar{\rho}\Delta_{\theta}\sin^{2}\theta}{\Xi^{2}}[adt-(r^{2}+a^{2})d\varphi]^{2}
-\frac{1}{\rho\bar{\rho}}(\frac{dr^{2}}{\Delta_{r}}+\frac{d\theta^{2}}{\Delta_{\theta}}),
\end{eqnarray}
where
\begin{eqnarray}\label{2}
&\rho=-\frac{1}{r-ia\cos\theta},\nonumber\\
&\Delta_{r}=(r^{2}+a^{2})(1-\frac{\Lambda}{3}r^{2})-2Mr+Q^{2},\nonumber\\
&\Delta_{\theta}=1+\frac{\Lambda}{3}a^{2}\cos^{2}\theta,\nonumber\\
&\Xi=1+\frac{\Lambda}{3}.
\end{eqnarray}
Here $\Lambda$ is the cosmological constant, $M$ is the mass of the black hole, $Q$ is its charge and a is the angular momentum per unit mass.

The event horizon equation for the spacetime defined by metric (1) is
\begin{eqnarray}\label{3}
\Delta_{r}&=(r^{2}+a^{2})(1-\frac{\Lambda}{3}r^{2})-2Mr+Q^{2} \nonumber\\
&=-\frac{\Lambda}{3}(r-r_{-1})(r-r_{0})(r-r_{1})(r-r_{2})=0.
\end{eqnarray}
For the case of $\Lambda>0$, Eq. (3) typically possesses three positive real roots [27]. Assuming these roots are $r_{0}$, $r_{1}$, and $r_{2}$, with $r_{2}>r_{1}>r_{0}$, then $r_{2}$ can be identified as the cosmological horizon, whereas $r_{1}$ and $r_{0}$ may be regarded as the outer and inner black hole horizons, respectively. The remaining root, $r_{-1}$, is negative and therefore unphysical.

By analogy with the angular velocity and surface gravity of a black hole horizon, we define corresponding quantities $\Omega_{j}$ and $\kappa_{j}$ for each root of Eq. (3). These are expressed as:
\begin{equation}\label{4}
\Omega_{j}=\frac{a}{r_{j}^{2}+a^{2}}, \quad (j=-1, 0, 1, 2),
\end{equation}
and
\begin{equation}\label{5}
\kappa_{j}=\frac{\Delta'_{r_{j}}}{2\Xi(r_{j}^{2}+a^{2})}, \quad (j=-1, 0, 1, 2),
\end{equation}
where the subscripts match those of the roots, and
\begin{equation}\label{6}
\Delta'_{r_{j}}=-\frac{\Lambda}{3}\Pi^{2}_{k=-1(k\neq j)}(r_{j}-r_{k}).
\end{equation}
The quantities $\Omega_{j}$ and $\kappa_{j}$ satisfy the following equations:
\begin{equation}\label{7}
\sum^{2}_{j=-1}\frac{\Omega_{j}}{\kappa_{j}}=0,
\end{equation}
and
\begin{equation}\label{8}
\sum^{2}_{j=-1}\frac{1}{\kappa_{j}}=0.
\end{equation}
Following our labeling convention, $\Omega_{1}$ and $\Omega_{2}$ represent the angular velocities of the black hole horizon and the cosmological horizon, respectively; likewise, $\kappa_{1}$ and $\kappa_{2}$ denote their respective surface gravities.  It should be noted that the Kerr-Newman-de Sitter metric is Petrov type-D [28].

Although spin field components are typically coupled in curved spacetimes, the perturbation equations for the massless scalar ($s=0$, $p=0$), Weyl neutrino ($s=1/2$, $p=\pm 1/2$), electromagnetic ($s=1$, $p=\pm 1$), Rarita-Schwinger ($s=3/2$, $p=\pm 3/2$), and gravitational ($s=2$, $p=\pm 2$) fields in a Petrov type-D spacetime not only decouple into pairs but also share a common structure, being governed by a single unified equation of the form [29]
\begin{equation}\label{9}
[(\nabla^{\mu}+pL^{\mu})(\nabla_{\mu}+pL_{\mu})-4p^{2}\psi_{2}+\frac{1}{6}R]\Phi_{p}=0.
\end{equation}
Here, the function $\Phi_{p}$ is related to the wave function $\chi^{(s)}_{p}$ of the spin field by
\begin{equation}\label{10}
\chi_{p}^{(s)}=H^{p-s}\Phi_{p},
\end{equation}
with $H$ is termed the generating function. The symbols $\nabla_{\mu}$, $L^{\mu}$, $\psi_{2}$, and $R$ in Eq. (9) represent the covariant derivative, spin-coefficient connection, Weyl scalar, and scalar curvature, respectively. For the Kerr-Newman-de Sitter metric, the explicit forms of these quantities, along with that of $H$, are given by [29]:
\begin{eqnarray}\label{11}
&L^{t}=-\rho\bar{\rho} \Xi[(r^{2}+a^{2})\frac{\Delta_{r}'}{2 \Delta_{r}}+\frac{2}{\bar{\rho}}+ia\sin\theta(\frac{\Delta_{\theta}'}{2 \Delta_{\theta}}+\cot\theta)],\nonumber\\
&L^{r}=-\frac{1}{2}\rho\bar{\rho} \Delta_{r}',\nonumber\\
&L^{\theta}=0,\nonumber\\
&L^{\varphi}=-\rho\bar{\rho}\Xi[a\frac{\Delta_{r}'}{2\Delta_{r}}+\frac{i}{\sin\theta}(\frac{\Delta_{\theta}'}{2\Delta_{\theta}}+\cot\theta)];
\end{eqnarray}
\begin{equation}\label{12}
\psi_{2}=\rho^{3}(M+\bar{\rho}Q^{2});
\end{equation}
\begin{equation}\label{13}
R=4\Lambda;
\end{equation}
and
\begin{equation}\label{14}
H=r-ia\cos\theta.
\end{equation}
where the prime denotes the derivative with respect to the independent variable.

Substituting Eqs. (11-13) into Eq. (9), we find that the resulting equation is separable. Its solution takes the form:
\begin{eqnarray}\label{15}
\Phi_{p}=\frac{e^{-i\omega t+im\varphi+i\omega r_{*}}}{(r-r_{-1})^{2p+1}}\exp[-im\sum^{2}_{j=0}\frac{\Omega_{j}}{2\kappa_{j}}\ln(\frac{r-r_{j}}{r-r_{-1}})]S_{p}(\theta)y_{p}(r),
\end{eqnarray}
where $r_{*}$ denotes the tortoise coordinate and is defined by the relation
\begin{equation}\label{16}
r_{*}=\int\frac{\Xi(r^{2}+a^{2})}{\Delta_{r}}dr=\sum^{2}_{j=-1}\frac{1}{2\kappa_{j}}\ln\mid\frac{r}{r_{j}}-1\mid.
\end{equation}
Note that as $r$ approaches the black hole horizon $r_{1}$, $r_{*}$ tends to $-\infty$, while it tends to $\infty$ as $r$ approaches the cosmological horizon $r_{2}$. For $\Phi_{p}$ to be single-valued, i.e., for $\Phi_{p}(\varphi+2\pi)=\Phi_{p}(\varphi)$ to hold, the parameter $m$ in Eq. (15) must be restricted to $m=0, \pm 1, \pm 2, ...$; this is both a necessary and sufficient condition.

It is convenient to introduce a new coordinate $z$ via the transformation
\begin{equation}\label{17}
z=\frac{(r_{2}-r_{-1})(r-r_{1})}{(r_{2}-r_{1})(r-r_{-1})},
\end{equation}
which maps the black hole horizon to $z=0$ and the cosmological horizon to $z=1$. The positivity of the denominator in Eq. (17) guarantees that this mapping is singularity-free.

Through substitution of Eqs. (11)-(13) and (15) into Eq. (9) and application of Eqs. (7), (8) and (17), it is found that $y_{p}$ satisfies:
\begin{eqnarray}\label{18}
\frac{d^{2}y_{p}}{dz^{2}}+(\frac{\gamma}{z}+\frac{\delta}{z-1}+\frac{\epsilon}{z-z_{0}})\frac{dy_{p}}{dz}+\frac{\alpha\beta z-q}{z(z-1)(z-z_{0})}y_{p}=0,
\end{eqnarray}
with
\begin{equation}\label{19}
\gamma+\delta+\epsilon=\alpha+\beta+1.
\end{equation}
Here,
\begin{eqnarray}\label{20}
&&\gamma=\frac{i}{\kappa_{1}}(\omega-m\Omega_{1})+p+1, \quad \delta=\frac{i}{\kappa_{2}}(\omega-m\Omega_{2})+p+1, \nonumber\\
&&\epsilon=\frac{i}{\kappa_{0}}(\omega-m\Omega_{0})+p+1, \quad z_{0}=z(r_{0})=\frac{(r_{2}-r_{-1})(r_{0}-r_{1})}{(r_{2}-r_{1})(r_{0}-r_{-1})}, \nonumber\\
&&\alpha=-\frac{i}{\kappa_{-1}}(\omega-m\Omega_{-1})+p+1, \quad \beta=2p+1,  \quad q=q(\lambda).
\end{eqnarray}
We omit the lengthy explicit expression for $q$, but note that it contains the separation constant $\lambda$. Its value will ultimately be determined by the quasinormal mode condition.

Equation (18) is a standard Heun equation [30] with regular singularities at $0, 1, z_{0}, \infty$ (with exponents $\{0, 1-\gamma\}$, $\{0, 1-\delta\}$, $\{0, 1-\epsilon\}$, $\{\alpha, \beta\}$, respectively) and with the following parameters: the singularity parameter $z_{0}$, the exponent parameters $\alpha$, $\beta$, $\gamma$, $\epsilon$, and the accessory parameter $q$. By relation (19), there are six free parameters in total.

The two local solutions of Eq. (18) at $z=0$, corresponding to its exponents, are given by [30]: the exponent $0$ yields
\begin{equation}\label{21}
H\ell(z_{0},q;\alpha,\beta,\gamma,\delta;z);
\end{equation}
and the exponent $1-\gamma$ yields
\begin{equation}\label{22}
z^{1-\gamma} H\ell(z_{0},(z_{0}\delta+\epsilon)(1-\gamma)+q;\alpha+1-\gamma,\beta+1-\gamma,2-\gamma,\delta;z).
\end{equation}

Provided that $\gamma \neq 0, -1, -2, \ldots$, the function $H\ell(z_{0},q;\alpha,\beta,\gamma,\delta;z)$ exists and is analytic in the disk $|z|<1$, with its Maclaurin expansion given by [30]
\begin{equation}\label{23}
H\ell(z_{0},q;\alpha,\beta,\gamma,\delta;z)=\sum^{\infty}_{j=0}c_{j}z^{j}, \quad |z|<1,
\end{equation}
where $c_{0}=1$,
\begin{eqnarray}\label{24}
&&z_{0}\gamma c_{1}=q c_{0}, \nonumber\\
&&z_{0}(j+1)(j+\gamma)c_{j+1}-j[(j-1+\gamma)(1+z_{0})+z_{0}\delta+\epsilon]c_{j} \nonumber\\
&&+(j-1+\alpha)(j-1+\beta)c_{j-1}=q c_{j}, \quad j\geq 1.
\end{eqnarray}

\section{Quasinormal mode Frequencies and radial wave functions}

Quasinormal modes are solutions to the wave equation defined by the boundary conditions of a purely ingoing wave at the event horizon and a purely outgoing wave at spatial infinity for asymptotically flat black holes. As an example, the asymptotic behavior of the radial wave function for the Kerr-Newman black hole is [31-33]
\begin{equation}\label{25}
R_{p}\sim\
\left\{
\begin{array}{ c c }
e^{-i(\omega-m\Omega_{1}) r_{\ast}}, & r_{\ast}\rightarrow -\infty, \quad (r\rightarrow r_{1});\\
e^{i\omega r_{\ast}}, & r_{\ast}\rightarrow \infty, \quad (r\rightarrow \infty). \\
\end{array}
\right.
\end{equation}

Spacetime (1) is asymptotically de Sitter. While its boundary condition at the event horizon remains the same as in an asymptotically flat black hole [4], the other boundary lies not at spatial infinity but at the cosmological horizon $r_{2}$. However, under the coordinate transformation (16), the cosmological horizon $r_{2}$ is mapped to infinity, rendering the coordinate $r_{*}$ range identical to that in the asymptotically flat case. It is therefore natural to postulate that for a Kerr-Newman-de Sitter black hole, the boundary condition at the cosmological horizon ($r_{*}\rightarrow \infty$) is a purely outgoing wave. This implies that, when the influence of the cosmological horizon is accounted for, the quasinormal mode boundary conditions can be formulated as
\begin{equation}\label{26}
R_{p}\sim\
\left\{
\begin{array}{ c c }
e^{-i(\omega-m\Omega_{1}) r_{\ast}}, & r_{\ast}\rightarrow -\infty, \quad (r\rightarrow r_{1});\\
e^{i(\omega-m\Omega_{2}) r_{\ast}}, & r_{\ast}\rightarrow \infty, \quad (r\rightarrow r_{2}). \\
\end{array}
\right.
\end{equation}

To find the radial wave function satisfying these conditions, we use Equation (18), the general solution of which is expressed as
\begin{eqnarray}\label{27}
y_{p}&=D_{1} H\ell(z_{0},q;\alpha,\beta,\gamma,\delta;z) \nonumber\\
&+D_{2} z^{1-\gamma} H\ell(z_{0},(z_{0}\delta+\epsilon)(1-\gamma)+q;\alpha+1-\gamma,\beta+1-\gamma,2-\gamma,\delta;z), \nonumber\\
\end{eqnarray}
where $D_{1}$ and $D_{2}$ are arbitrary constants. Then equation (15) takes the form
\begin{eqnarray}\label{28}
\Phi_{p}&=\frac{e^{-i\omega t+im\varphi+i\omega r_{*}}}{(r-r_{-1})^{2p+1}}\exp[-im\sum^{2}_{j=0}\frac{\Omega_{j}}{2\kappa_{j}}\ln(\frac{r-r_{j}}{r-r_{-1}})]S_{p}(\theta)\nonumber\\
&\cdot[D_{1} H\ell(z_{0},q;\alpha,\beta,\gamma,\delta;z) \nonumber\\
&+D_{2} z^{1-\gamma} H\ell(z_{0},(z_{0}\delta+\epsilon)(1-\gamma)+q;\alpha+1-\gamma,\beta+1-\gamma,2-\gamma,\delta;z)]. \nonumber\\
\end{eqnarray}
Hence, the radial wave function $R_{p}$ for massless particles of arbitrary spin is given by
\begin{eqnarray}\label{29}
R_{p}&=\frac{e^{i\omega r_{*}}}{(r-r_{-1})^{2p+1}}\exp[-im\sum^{2}_{j=0}\frac{\Omega_{j}}{2\kappa_{j}}\ln(\frac{r-r_{j}}{r-r_{-1}})][D_{1} H\ell(z_{0},q;\alpha,\beta,\gamma,\delta;z) \nonumber\\
&+D_{2} z^{1-\gamma} H\ell(z_{0},(z_{0}\delta+\epsilon)(1-\gamma)+q;\alpha+1-\gamma,\beta+1-\gamma,2-\gamma,\delta;z)]. \nonumber\\
\end{eqnarray}

Near the event horizon, where $r_{*}\rightarrow -\infty$ ($r\rightarrow r_{1}$), the radial wave function takes the asymptotic form:
\begin{equation}\label{30}
R_{p}=D_{1} e^{i (\omega-m\Omega_{1}) r_{*}}+D_{2} e^{-[2\kappa_{1} p+i (\omega-m\Omega_{1})]r_{*}}.
\end{equation}
Applying the boundary condition at $r=r_{1}$, which requires $D_{1}=0$, yields
\begin{eqnarray}\label{31}
R_{p}=&D_{2}\frac{z^{1-\gamma} e^{i\omega r_{*}}}{(r-r_{-1})^{2p+1}}\exp[-im\sum^{2}_{j=0}\frac{\Omega_{j}}{2\kappa_{j}}\ln(\frac{r-r_{j}}{r-r_{-1}})] \nonumber\\
& H\ell(z_{0},(z_{0}\delta+\epsilon)(1-\gamma)+q;\alpha+1-\gamma,\beta+1-\gamma,2-\gamma,\delta;z).
\end{eqnarray}

To obtain the quasinormal modes, we examine the asymptotic behavior of the radial wave function (31) as $r_{*} \rightarrow \infty$, which corresponds to $r \rightarrow r_{2}$ and $z \rightarrow 1$:
\begin{equation}\label{32}
R_{p}\sim e^{i (\omega-m\Omega_{2}) r_{*}} H\ell[(z_{0},(z_{0}\delta+\epsilon)(1-\gamma)+q;\alpha+1-\gamma,\beta+1-\gamma,2-\gamma,\delta;z)].
\end{equation}

Following the reasoning applied to Equation (23), if $2-\gamma \neq 0, -1, -2, ...$, the Heun function in Eq. (32) exists and is analytic within the disk $\mid z \mid <1$, where it admits a Maclaurin series expansion. Nevertheless, this analyticity does not ensure the convergence of the series at $z=1$. Consequently, the radial wave function (31) fails to satisfy the boundary condition at infinity.

The only way to resolve this is to force the series for the quasinormal modes to truncate, turning it into a Heun polynomial. We thus obtain
\begin{equation}\label{33}
H\ell[(z_{0},(z_{0}\delta+\epsilon)(1-\gamma)+q;\alpha+1-\gamma,\beta+1-\gamma,2-\gamma,\delta;1)]=finite\,value.
\end{equation}
Therefore, the radial wave function meets the boundary condition required for quasinormal modes as $r_{*} \to \infty$.

The condition that the Heun series in Eq. (31) becomes a polynomial of degree $n$ is
\begin{equation}\label{34}
\alpha+1-\gamma==-n, \quad (n=0, 1, 2, ...),
\end{equation}
This yields an expression for the quasinormal mode frequencies:
\begin{equation}\label{35}
\omega=[-i(n+1)+m(\frac{\Omega_{1}}{\kappa_{1}}+\frac{\Omega_{-1}}{\kappa_{-1}})](\frac{1}{\kappa_{1}}+\frac{1}{\kappa_{-1}})^{-1}.
\end{equation}
Equation (35) shows that all massless particles with spin $\leq 2$ can share the same quasinormal mode frequencies for given quantum numbers $n$ and $m$. Since these frequencies depend only on the black hole's properties and not on the particles themselves, one type of particle can mimic the quasinormal modes of another.

The boundary conditions for quasinormal modes require that the Heun function in Eq. (31) be a polynomial. For a polynomial of degree $n$, the coefficients $c_j$ from the recurrence relation (24) must satisfy the following matrix equation:
\begin{equation}\label{36}
\begin{array}{ll}
\left(\begin{array}{lll}
\,\,0\,\,\,\,\,\,\,\,\,\,\,\,\,C_{0}\,\,\,\,\,\,\,\,\,\,\,\,\,\,0\,\,\,\,\,\,\,\,\,\cdots\,\,\,\,\,\,\,\,0 \\
A_{1}\,\,\,\,\,\,-B_{1}\,\,\,\,\,\,\,\,\,\,C_{1}\,\,\,\,\,\,\,\cdots\,\,\,\,\,\,\,\,0 \\
\,\,0\,\,\,\,\,\,\,\,\,\,\,\,\,A_{2}\,\,\,\,\,\,-B_{2}\,\,\,\,\,\,\,\,\,\,\,\,\,\,\,\,\,\,\,\,\,\,\,\,\vdots \\
\,\,\,\vdots\,\,\,\,\,\,\,\,\,\,\,\,\,\,\,\,\vdots\,\,\,\,\,\,\,\,\,\,\,\,\,\,\,\,\,\,\vdots\,\,\,\,\,\,\,\,\,\,\ddots\,\,\,\,\,C_{n-1} \\
\,\,0\,\,\,\,\,\,\,\,\,\,\,\,\,\,\,0\,\,\,\,\,\,\,\,\,\,\,\,\,\,\cdots\,\,\,\,\,\,\,A_{n}\,\,\,\,-B_{n}
\end{array}
\right)
\left(\begin{array}{lll}
\,\,\,\,c_{0}\\
\,\,\,\,c_{1}\\
\,\,\,\,\,\vdots \\
c_{n-1}\\
\,\,\,\,c_{n}
\end{array}
\right)
=Q\left(\begin{array}{lll}
\,\,\,\,c_{0}\\
\,\,\,\,c_{1}\\
\,\,\,\,\,\vdots \\
c_{n-1}\\
\,\,\,\,c_{n}
\end{array}
\right),
\end{array}
\end{equation}
with $c_{0}=1$,
\begin{eqnarray}\label{37}
A_{j}&=(j+\alpha-\gamma)(j+\beta-\gamma) \nonumber\\
&=(j-1-n)(j+p-\frac{(n+1)\kappa_{-1}}{\kappa_{1}+\kappa_{-1}}+\frac{im(\Omega_{1}-\Omega_{-1})}{\kappa_{1}+\kappa_{-1}}), \nonumber\\
B_{j}&=j[(j+1-\gamma)(1+z_{0})+z_{0}\delta+\epsilon] \nonumber\\
&=j[(1+z_{0})(j-\frac{(n+1)\kappa_{-1}}{\kappa_{1}+\kappa_{-1}}+\frac{im(\Omega_{1}-\Omega_{-1})}{\kappa_{1}+\kappa_{-1}}) \nonumber\\
&+z_{0}(1-\frac{(n+1)\kappa_{0}}{\kappa_{0}+\kappa_{2}}+\frac{im(\Omega_{0}-\Omega_{2})}{\kappa_{0}+\kappa_{2}}) \nonumber\\
&+1-\frac{(n+1)\kappa_{2}}{\kappa_{0}+\kappa_{2}}+\frac{im(\Omega_{2}-\Omega_{0})}{\kappa_{0}+\kappa_{2}}] \nonumber\\
C_{j}&=z_{0}(j+1)(j+2-\gamma) \nonumber\\
&=z_{0}(j+1)(j+1-p-\frac{(n+1)\kappa_{-1}}{\kappa_{1}+\kappa_{-1}}+\frac{im(\Omega_{1}-\Omega_{-1})}{\kappa_{1}+\kappa_{-1}}), \nonumber\\
Q&=(z_{0}\delta+\epsilon)(1-\gamma)+q.
\end{eqnarray}

Algebraic theory states that a necessary condition for a non-trivial solution is for $Q$ to be an eigenvalue of the tridiagonal matrix in Eq. (36), which constrains $Q$ to the discrete values $Q = Q_{n,k}$ for $k = 0, 1, 2, \dots, n$. We postulate that the eigenvector corresponding to $Q_{n,k}$ is $(c^{0}_{n,k}\,\,c^{1}_{n,k}\,\,...\,\,c^{n}_{n,k})^{T}$. The final form of the quasinormal-mode radial wave function is therefore
\begin{eqnarray}\label{38}
R_{p}&=D_{2}\frac{ z^{1-\gamma}e^{i\omega r_{*}}}{(r-r_{-1})^{2p+1}}\exp[-im\sum^{2}_{j=0}\frac{\Omega_{j}}{2\kappa_{j}}\ln(\frac{r-r_{j}}{r-r_{-1}})]  \nonumber\\
&\cdot H\ell(a,Q_{n,k};-n,\beta+1-\gamma,2-\gamma,\delta;z) \nonumber\\
&=D_{2}\frac{ z^{1-\gamma}e^{i\omega r_{*}}}{(r-r_{-1})^{2p+1}}\exp[-im\sum^{2}_{j=0}\frac{\Omega_{j}}{2\kappa_{j}}\ln(\frac{r-r_{j}}{r-r_{-1}})] \sum^{n}_{j=0}c^{j}_{n,k} z^{j}.
\end{eqnarray}

\section{ Discussion and conclusion}
In this paper, we have obtained analytical formulas for the quasinormal modes of Kerr-Newman-de Sitter black holes. The quasinormal mode frequency (35) is complex-valued, which physically corresponds to underdamped oscillatory behavior. The real part of the frequency, representing the actual oscillation, is proportional to the quantum number $m$. Conversely, the imaginary part, which determines the damping rate, is proportional to the quantum number $n+1$. Notably, the oscillation frequency varies discretely and equally with $m$, while the damping rate varies discretely and equally with $n$. These quasinormal modes constitute a countable set of discrete frequencies, demonstrating that black holes share characteristics analogous to ``harmonic oscillators'' in gravitational physics.

The quasinormal mode frequencies are independent of the spin of the perturbing fields, being determined solely by the black hole's parameters. For given quantum numbers $n$ and $m$, the quasinormal mode frequency is identical for massless particles of any spin. This universality originates from the structure of the Kerr-Newman-de Sitter spacetime, which is bounded by two horizons--the event horizon and the cosmological horizon. This implies that the gravitational ``fingerprints'' of a black hole are identical for all types of massless spin particles. Consequently, through the observation of quasinormal modes, one can not only determine the black hole's parameters but also witness the phenomenon where one type of particle perfectly mimics the quasinormal modes of another. This study thus provides a theoretical foundation for this mimicking phenomenon.

The quasinormal mode frequency (35) depends only on the quantum numbers $n$ and $m$, but not on $k$. However, the radial wave function (38) depends on all three quantum numbers $n$, $m$, and $k$, implying that the quasinormal mode frequency is degenerate. Given that $k$ takes values $k = 0, 1, 2, ..., n$, the degree of degeneracy is $n+1$. Analytical solutions for quasinormal modes are extremely rare, and without such expressions, discussing the degeneracy of their frequencies is difficult. This explains why very few studies in the literature have addressed this issue. Therefore, our study is of significant importance for a comprehensive understanding of the full characteristics of quasinormal modes.

\ack
This work was supported by the National Natural Science Foundation
of China under Grant No. 12175198.


\section*{References}

\begin{thebibliography}{99}

\bibitem{1} BERTI E., CARDOSO V. and STARINETS A. O.,\textsl{ Class. Quant. Grav}., {\bf 26} (2009) 163001.

\bibitem{2} VISHVESHWARA C., \textsl{Nature}, {\bf 227} (1970) 936.

\bibitem{3} KOKKOTAS K. D. and SCHMIDT B. G., \textsl{Living Rev. Rel.}, {\bf 2}(1999) 2.

\bibitem{4} KONOPLYA R. A. and ZHIDENKO A., \textsl{Rev. Mod. Phys.}, {\bf 83} (2011) 793.

\bibitem{5} MOULIN F., BARRAU A. and MARTINEAU K., \textsl{Universe}, {\bf 5} (2019) 202.

\bibitem{6} BOLOKHOV S. V. and SKVORTSOVA M., \textsl{Review of analytic results on quasinormal modes of black holes}, arXiv:2504.05014.

\bibitem{7} SHEN S.-F., LI G.-R., DAGHIGH R. G., MOREY J. C., GREEN M. D., QIAN W.-L. and YUE R.-H., \textsl{Asymptotic quasinormal modes, echoes, and black hole spectral instability: a brief review}, arXiv:2507.11663.

\bibitem{8} ABBOTT B.P. et al. (LIGO Scientific, Virgo), \textsl{Phys. Rev. Lett.}, {\bf 116} (2016) 061102.

\bibitem{9} BOURG P., MACEDO R. P., SPIERS A., LEATHER B., BONGA B. and POUND A., \textsl{Phys. Rev. Lett.}, {\bf 134} (2025) 061401.


\bibitem{10} OSHITA N., BERTI E. and CARDOSO V., \textsl{Phys. Rev. Lett.}, {\bf 135} (2025) 031401.

\bibitem{11} ZHONG Z., CARDOSO V. and CHEN Y.,\textsl{ Phys. Rev. Lett.}, {\bf 134} (2025) 211402.

\bibitem{12} BURGESS C., PATRICK S., TORRES T., GREGORY R. and K$\ddot{O}$NIG F., \textsl{Phys. Rev. Lett.}, {\bf 132} (2024) 053802.

\bibitem{13} SOLIDORO L., PATRICK S., WEINFURTNER S. and GREGORY R., \textsl{Phys. Rev. Lett.}, {\bf 135} (2025) 051401.

\bibitem{14} LESCLUZE M DE. and HELLER M. P., \textsl{Phys. Rev. Lett.}, {\bf 135} (2025) 091601.

\bibitem{15} HOLZM$\ddot{U}$LLER G., \textsl{Zeitschrift f$\ddot{u}$r Mathematik und Physik}, {\bf 15} (1870) 69.

\bibitem{16} TISSERAND F., \textsl{Comptes Rendus de l' Acad$\grave{e}$mie des Sciences (Paris)}, {\bf 75} (1872) 760.

\bibitem{17} TISSERAND F., \textsl{Comptes Rendus de l' Acad$\grave{e}$mie des Sciences (Paris)}, {\bf 100} (1890) 313.

\bibitem{18} BOYENENI S., WU J. and MOST E. R., \textsl{Phys. Rev. Lett.}, {\bf 135} (2025) 101401.

\bibitem{19} TEUKOLSKY S. A., \textsl{Phys. Rev. Lett.}, {\bf 29} (1972) 1114.

\bibitem{20} TEUKOLSKY S. A., \textsl{Astrophys. J.}, {\bf 185} (1973) 635.

\bibitem{21} DANZMANN K., \textsl{Class. Quant. Grav.}, {\bf 14} (1997) 1399.

\bibitem{22} LISA collaboration, \textsl{Laser interferometer space antenna}, arXiv:1702.00786.

\bibitem{23} HU W.-R. and WU Y.-L., \textsl{Natl. Sci. Rev.}, {\bf 4} (2017) 685.

\bibitem{24} TianQin collaboration, \textsl{Class. Quant. Grav.}, {\bf 33} (2016) 035010.

\bibitem{25} CARTER B., \textsl{Commun. Math. Phys.}, {\bf 17} (1970) 233.

\bibitem{26} GIBBONS G. W. and HAWKING S.W.,\textsl{ Phys. Rev. D}, {\bf 15} (1977) 2738.

\bibitem{27} JIANG M., WANG X. and LI Z.-H., \textsl{Astrophys. Space Sci.}, {\bf 357} (2015) 139.

\bibitem{28} LI Z.-H., \textsl{Phy. Lett. B}, {\bf 643} (2006) 64.

\bibitem{29} LI Z.-H., \textsl{Unified equation for massless spin particles and new spin coefficient definitions}, arXiv:2504.02592.

\bibitem{30} RONVEAUX A., \textsl{Heun's Differential Equations}, 1st edn. (Clarendon Press, Oxford, 1995).

\bibitem{31} Kokkotas, K. D., \textsl{Nuovo Cimento Soc. Ital. Fis. B}, {\bf 108} (1993) 991.

\bibitem{32} BERTI E. and KOKKOTAS K. D., \textsl{Phys. Rev. D}, {\bf 71} (2005) 124008.

\bibitem{33} JING J. and PAN Q., \textsl{Nucl. Phys. B}, {\bf 728} (2005) 109.


\end{thebibliography}
\end{document}